\documentclass[11pt,preprint,letterpaper]{aastex}

\usepackage{amssymb,amsmath}

\usepackage[squaren, textstyle]{SIunits}
\usepackage{url}
\usepackage{natbib}
\usepackage{longtable}
\usepackage{lastpage}

\newcommand{\pV}{\ensuremath{p_\text{V}{}}}
\newcommand{\pv}{\pV}

\newcommand{\Jy}{\ensuremath{\text{Jy}}}

\newcommand{\jy}{\Jy}

\newcommand{\km}{\kilo\metre}

 \newcommand{\dv}{\ensuremath{\Delta v}}

\newcommand{\Delbo}{Delbo'}

\newcommand{\Fernandez}{Fern\'{a}ndez} 

\newcommand{\Juric}{Juri\'c}

\newcommand{\Michalowski}{Micha\l owski}

\newcommand{\Sarounova}{\v{S}arounov\'{a}}

\newcommand{\revision}[1]{#1}
\slugcomment{Submitted to \aj\ on 2010/09/25, accepted 2011/01/25}
\shorttitle{ExploreNEOs paper III: Low-deltaV objects}
\shortauthors{Mueller et al.}

\begin{document}

  \title{ExploreNEOs III: Physical characterization of 65 low-deltaV NEOs}

\author{Michael Mueller\altaffilmark{1}}
\author{M.\ \Delbo\altaffilmark{1}}
\author{J.L.\ Hora\altaffilmark{2}}
\author{D.E.\ Trilling\altaffilmark{3}}
\author{B.\ Bhattacharya\altaffilmark{4}}
\author{W.F.\ Bottke\altaffilmark{5}}
\author{S.\ Chesley\altaffilmark{6}}
\author{J.P.\ Emery\altaffilmark{7}}
\author{G.\ Fazio\altaffilmark{2}}
\author{A.W.\ Harris\altaffilmark{8}}
\author{A.\ Mainzer\altaffilmark{6}}
\author{M.\ Mommert\altaffilmark{8}}
\author{B.\ Penprase\altaffilmark{9}}
\author{H.A.\ Smith\altaffilmark{2}}
\author{T.B.\ Spahr\altaffilmark{2}}
\author{J.A.\ Stansberry\altaffilmark{10}}
\and
\author{C.A.\ Thomas\altaffilmark{3}}

\altaffiltext{1}{UNS-CNRS-Observatoire de la C\^ote d'Azur, 
Laboratoire Cassiop\'ee, 
BP 4229, 
06304 Nice cedex 04, France; 
\email{michael.mueller@oca.eu}}
\altaffiltext{2}{Harvard-Smithsonian Center for Astrophysics, 
60 Garden St., MS-65, 
Cambridge, MA 02138}
\altaffiltext{3}{Department of Physics and Astronomy,
Northern Arizona University, Flagstaff, AZ 86001}
\altaffiltext{4}{NASA Herschel Science Center, Caltech, M/S 100-22, 770 South Wilson Ave.
Pasadena, CA  91125  USA}
\altaffiltext{5}{Southwest Research Institute,  1050 Walnut St, Suite 300, Boulder, Colorado 80302, USA}
\altaffiltext{6}{Jet Propulsion Laboratory, California Institute of Technology, Pasadena, CA 91109, USA}
\altaffiltext{7}{Department of Earth and Planetary Sciences,
University of Tennessee,
1412 Circle Dr.,
Knoxville, TN 37996}
\altaffiltext{8}{DLR Institute of
  Planetary Research, Rutherfordstrasse 2, 12489 Berlin, Germany.}
\altaffiltext{9}{Department of Physics and
Astronomy, Pomona College, 610 N. College Ave,
Claremont, CA 91711}
\altaffiltext{10}{Steward Observatory,
University of Arizona,
933 N. Cherry Ave,
Tucson AZ 85721}

\begin{abstract}

Space missions to NEOs are being planned at all major space agencies, 
and recently a manned mission to an NEO was announced as a NASA goal.
Efforts to find
and select suitable targets (plus backup targets) are severely hampered by our lack of
knowledge 
\revision{of}
the physical properties of dynamically favorable NEOs. In particular,
current mission scenarios tend to favor primitive low-albedo objects. For the vast
majority of NEOs the albedo is unknown.

Here we report new constraints on the size and albedo of 65 NEOs with rendezvous deltaV $< 7$
km/s. Our results are based on thermal-IR flux data obtained in the framework of our
ongoing (2009--2011) ExploreNEOs survey \citep{Trilling2010} using NASA's ``Warm Spitzer'' space telescope. 
As of 2010 July 14, we have results for 293 objects in hand (including the 65
low-deltaV NEOs presented here); before the end of 2011 we expect to have measured
the size and albedo of $\sim 700$ NEOs (including probably $\sim 160$ low-deltaV NEOs). 

While there are reasons to believe that primitive volatile-rich materials are universally
low in albedo, the converse need not be true: 
\revision{t}he
orbital evolution of some dark objects likely has
caused them to lose their volatiles by coming too close to the Sun. For all our targets,
we give the closest perihelion distance they are likely to have reached 
\citep[using orbital integrations from][]{Marchi:2009p5116}
and corresponding upper limits on the past
surface temperature.
Low-deltaV objects for which both albedo and thermal history \revision{may} suggest a primitive
composition include (162998) 2001 SK162, (68372) 2001 PM9, and (100085) 1992 UY4. 
\end{abstract}



\section{Introduction}

To date, two near-Earth objects (NEOs) have been targeted by 
space missions, both yielding a wealth of fascinating and groundbreaking insights into the past and current state of the Solar System: NASA's Near-Shoemaker mission went into orbit around its target (433) Eros in 2000 and landed on it in the following year; 
\revision{the}
Japanese mission Hayabusa arrived at (25143) Itokawa in 2005 and scrutinized the NEO for a few months.  In June 2010, Hayabusa succeeded in returning 
asteroid dust samples to Earth.

Given the remarkable success of these missions 
it is perhaps not surprising that robotic NEO mission concepts are being considered at space agencies across the planet, including NASA's OSIRIS-REx, 
JAXA's Hayabusa 2, and ESA's Marco Polo \citep[see, e.g.,][for recent updates on these missions]{osiris,Hayabusa2,marcopolo}.
In a speech in April 2010, President Obama announced the goal of a manned space mission to an asteroid
\citep[see][for a corresponding NASA mission scenario including robotic precursor missions]{Abell2009}.

Finding a suitable target asteroid is one of the challenging aspects of mission planning. Targets are  tightly constrained in terms of their orbital dynamics and physical properties. Furthermore, launch windows are usually tight and the planning process long.  Unforeseen delays due to technological or financial problems risk eliminating the nominal target; it is therefore generally advisable to plan for contingency or backup targets.

\paragraph{Dynamics, \dv:}
As discussed by \citet{ShoemakerHelin1978}, mission cost depends chiefly on the required amount of propellant, which follows from the total specific linear momentum, \dv, that must be imparted on the spacecraft for it to reach the target orbit. Minimizing \dv\ is therefore a top priority for practical reasons.
It is worth emphasizing that there are a large number of NEOs that are reachable at a lower \dv\ than 
\revision{that required to reach}
Mars.

A realistic assessment of \dv\  depends on the specific mission scenario and timing, and must be evaluated on a target-by-target basis. 
\revision{A customary first-order estimate is the \dv\ of a Hohmann transfer orbit, which is an analytic function of the orbital elements \citep[see][]{ShoemakerHelin1978}.}
%
Thus, in a first target selection process objects with sufficiently small 
``Hohmann-\dv'' are identified. Only those objects need to be studied in detail.  For the purposes of this study, we will refer to the ``Hohmann-\dv'' as \dv\ (without qualifiers).

\paragraph{Physical properties:} 
For the vast majority of NEOs, including low-\dv\ objects, nothing is known about their physical properties.  
Frequently, however, mission concepts require the target to be within a given size or mass range, e.g., in order to enable the spacecraft to orbit the target.
Moreover, the science goals of some current  mission scenarios require their target to be a ``primitive'' object, translating into constraints on their albedo and thermal history (see below).

\paragraph{Primitive objects:}
Some 
\revision{meteorites} 
contain surprisingly pristine material
\revision{that}
has suffered very little modification since the early days of the Solar System.  
\revision{Their asteroidal parent bodies}
are of particular interest for some NEO missions (especially since both NEAR-Shoemaker and Hayabusa targeted S-type asteroids, which have undergone significant processing).

Judging from meteorite analogues, asteroids with very low albedo (geometric albedo $\pv \lesssim \unit{7.5}{\%}$) 
\revision{are very likely to}
be ``primitive'' and vice-versa  \citep{Fernandez2005}.
A word of caution applies to NEOs, however: their relative proximity to the Sun can potentially cause their surfaces to heat up to the point that thermal surface alterations occur (de-volatilization, chemical reactions, etc.).  
There are hence two necessary conditions for an NEO to have a primitive surface: a low albedo and an orbital history that never brought the perihelion too close to the Sun.  

\paragraph{ExploreNEOs:}
This is the third paper describing results from the ongoing (2009--2011) ExploreNEOs survey \citep[following][]{Trilling2010,Harris2011}.
The primary goal of this survey is to measure the size and albedo of $\sim700$ NEOs based on observations with NASA's ``Warm Spitzer'' space telescope.  \citet{Trilling2010} describe the goals and methods of the survey along with results for the first 101 NEOs; \citet{Harris2011} check the accuracy of 
\revision{the ExploreNEOs}
results against values published in the literature (where available) and find  diameters to be \revision{typically} consistent within 
\revision{20}\%,
albedos within 50\%.

As of 2010 July 14, we have data for 293 NEOs in hand including 65 objects with $\dv<\unit{7}{\kilo\metre\per\second}$. By the end of the survey, i.e.\ before the end of 2011, we expect to have measured $\sim 160$ low-\dv\ objects.
We chose to publish this first batch of results in order to alert the community to the existence of our growing database of characterized low-\dv\ objects.  


\paragraph{Overview of this work:}

In Section \ref{sect:obs} we describe our photometric data.  Our modeling approach is described in Section \ref{sect:thermal}, 
\revision{and}
resulting diameter and albedo estimates are presented in Section \ref{sect:dpv}.  In Section \ref{sect:history} we study the thermal history of our targets. We discuss the implications of our results in Section \ref{sect:discu} and summarize our conclusions in Section 
\ref{sect:conclu}.

\section{Warm-Spitzer observations}
\label{sect:obs}

The observations reported herein 
use the post-cryogenic (``warm'') mode of the IRAC camera \citep{IRAC} onboard the Spitzer Space Telescope \citep{SST}. 
Each NEO target is observed  in the two photometric channels (channel\revision{s} 1 and 2) with central wavelengths of around 3.6 and \unit{4.5}{\micron}, respectively. Observations
\revision{are built up from frames that}
 alternate repeatedly between the two channels, such that the resulting fluxes are quasi-simultaneous.
Further details on our observation design and data reduction are given in \citet{Trilling2010}.

\begin{deluxetable}{llllllllllll}
\tabletypesize{\scriptsize}
\tablecaption{\label{tab:fluxes} Spitzer data sorted by \dv}
\tablehead{
\dv & Object & Time & $H$ & $r$ & $\Delta$ & $\alpha$ & f36 & f45 \\
(\kilo\meter\per\second) &  &(UT)  & & (AU) & (AU) & (deg.) & (\milli\jy)  & (\milli\jy)  
}
\startdata
 4.632 & (25143) Itokawa & 2010-May-15 14:45:09 & 19.20 & 1.018 & 0.052 & 74.56 & $ 1.767 \pm 0.041 $ & $ 6.033 \pm 0.072 $ \\
 4.755 & 1996 XB27 & 2010-Jul-12 06:41:36 & 21.84 & 1.121 & 0.192 & 51.92 & $ 0.0178 \pm 0.0040 $ & $ 0.0428 \pm 0.0061 $ \\
 4.887 & (10302) 1989 ML & 2009-Aug-21 03:09:35 & 19.50 & 1.100 & 0.152 & 56.01 & $ 0.229 \pm 0.015 $ & $ 0.560 \pm 0.022 $ \\
 5.276 & (99799) 2002 LJ3 & 2009-Sep-19 22:38:44 & 18.10 & 1.238 & 0.354 & 46.18 & $ 0.125 \pm 0.015 $ & $ 0.392 \pm 0.020 $ \\
 5.280 & 2001 CQ36 & 2010-Apr-15 13:27:01 & 22.45 & 1.069 & 0.125 & 55.50 & $ 0.0215 \pm 0.0044 $ & $ 0.0787 \pm 0.0081 $ \\
 5.302 & (52381) 1993 HA & 2009-Nov-18 12:09:41 & 20.20 & 1.257 & 0.402 & 46.59 & $ 0.0303 \pm 0.0052 $ & $ 0.150 \pm 0.011 $ \\
 5.328 & 2000 YF29 & 2010-Feb-20 04:43:34 & 20.16 & 1.015 & 0.123 & 83.03 & $ 0.131 \pm 0.012 $ & $ 0.467 \pm 0.020 $ \\
 5.391 & (1943) Anteros & 2009-Sep-15 00:09:47 & 15.75 & 1.548 & 0.951 & 40.17 & $ 0.189 \pm 0.013 $ & $ 0.582 \pm 0.022 $ \\
 5.486 & (138911) 2001 AE2 & 2009-Aug-13 14:53:36 & 19.10 & 1.330 & 0.483 & 41.78 & $ 0.0244 \pm 0.0047 $ & $ 0.0888 \pm 0.0086 $ \\
 5.487 & 2006 SY5 & 2009-Jul-30 18:18:27 & 22.08 & 1.093 & 0.136 & 54.07 & $ 0.0312 \pm 0.0051 $ & $ 0.1167 \pm 0.0097 $ \\
 5.555 & (162416) 2000 EH26 & 2010-Apr-29 07:39:49 & 21.70 & 1.125 & 0.217 & 51.21 & $ 0.0378 \pm 0.0058 $ & $ 0.1181 \pm 0.0099 $ \\
 5.565 & (162998) 2001 SK162 & 2009-Dec-17 23:35:37 & 18.00 & 1.135 & 0.451 & 63.68 & $ 0.815 \pm 0.028 $ & $ 3.184 \pm 0.051 $ \\
 5.653 & (68372) 2001 PM9 & 2009-Aug-20 10:32:54 & 18.90 & 1.130 & 0.204 & 53.62 & $ 3.928 \pm 0.059 $ & $ 19.61 \pm 0.14 $ \\
 5.719 & (12923) Zephyr & 2010-May-21 04:54:42 & 16.10 & 1.477 & 0.876 & 41.50 & $ 0.253 \pm 0.021 $ & $ 0.453 \pm 0.023 $ \\
 6.041 & (85938) 1999 DJ4 & 2009-Aug-13 15:40:17 & 18.60 & 1.409 & 0.688 & 43.10 & $ 0.0172 \pm 0.0039 $ & $ 0.0620 \pm 0.0072 $ \\
 6.069 & (433) Eros & 2009-Aug-25 06:16:47 & 11.16 & 1.702 & 1.218 & 36.50 & $ 13.99 \pm 0.11 $ & $ 40.23 \pm 0.20 $ \\
 6.070 & 2000 GV147 & 2009-Oct-24 06:09:17 & 19.03 & 1.064 & 0.336 & 74.34 & $ 0.0998 \pm 0.010 $ & $ 0.337 \pm 0.017 $ \\
 6.086 & 2000 XK44 & 2010-Jan-11 03:12:17 & 17.73 & 1.288 & 0.848 & 51.85 & $ 0.0285 \pm 0.0051 $ & $ 0.1040 \pm 0.0094 $ \\
 6.106 & 1993 RA & 2009-Nov-17 02:57:14 & 18.91 & 1.191 & 0.587 & 59.20 & $ 0.0165 \pm 0.0039 $ & $ 0.0544 \pm 0.0068 $ \\
 6.130 & (177614) 2004 HK33 & 2009-Jul-30 16:25:59 & 17.60 & 1.057 & 0.095 & 63.88 & $ 5.895 \pm 0.071 $ & $ 22.92 \pm 0.14 $ \\
 6.191 & 1998 VO & 2009-Dec-13 03:44:37 & 20.37 & 1.024 & 0.021 & 76.62 & $ 4.481 \pm 0.063 $ & $ 15.31 \pm 0.11 $ \\
 6.196 & 2006 SV19 & 2009-Sep-15 16:08:48 & 17.76 & 1.166 & 0.672 & 60.97 & $ 0.109 \pm 0.010 $ & $ 0.435 \pm 0.019 $ \\
 6.236 & 2005 JA22 & 2009-Nov-10 03:50:52 & 18.47 & 1.296 & 0.489 & 46.66 & $ 0.0908 \pm 0.0093 $ & $ 0.319 \pm 0.017 $ \\
 6.240 & (1627) Ivar & 2010-Jun-16 08:39:40 & 13.20 & 1.933 & 1.249 & 27.62 & $ 1.264 \pm 0.033 $ & $ 2.670 \pm 0.048 $ \\
 6.279 & (87024) 2000 JS66 & 2010-Jan-10 19:24:49 & 18.70 & 1.104 & 0.567 & 65.82 & $ 0.0163 \pm 0.0023 $ & $ 0.0427 \pm 0.0061 $ \\
 6.323 & (22099) 2000 EX106 & 2010-Jan-11 09:47:46 & 18.00 & 1.029 & 0.246 & 79.60 & $ 0.225 \pm 0.015 $ & $ 0.842 \pm 0.027 $ \\
 6.364 & 2003 SL5 & 2009-Sep-19 23:06:19 & 19.14 & 1.114 & 0.164 & 53.62 & $ 0.289 \pm 0.017 $ & $ 1.052 \pm 0.029 $ \\
 6.364 & (35107) 1991 VH & 2010-Mar-20 17:04:49 & 16.90 & 1.172 & 0.660 & 58.78 & $ 0.113 \pm 0.011 $ & $ 0.477 \pm 0.021 $ \\
 6.379 & (172974) 2005 YW55 & 2010-Mar-24 09:16:52 & 19.30 & 1.242 & 0.386 & 44.41 & $ 0.0615 \pm 0.0078 $ & $ 0.171 \pm 0.013 $ \\
 6.405 & (65679) 1989 UQ & 2009-Oct-15 20:50:49 & 19.40 & 1.129 & 0.237 & 58.64 & $ 0.405 \pm 0.061 $ & $ 2.264 \pm 0.043 $ \\
 6.431 & (159402) 1999 AP10 & 2009-Aug-31 22:54:55 & 16.40 & 1.292 & 0.838 & 52.32 & $ 0.090 \pm 0.010 $ & $ 0.278 \pm 0.016 $ \\
 6.491 & (143651) 2003 QO104 & 2009-Aug-21 02:40:17 & 16.00 & 1.402 & 0.801 & 45.93 & $ 0.276 \pm 0.042 $ & $ 1.052 \pm 0.031 $ \\
 6.507 & 1989 AZ & 2010-Feb-19 07:05:07 & 19.49 & 1.003 & 0.044 & 93.49 & $ 8.998 \pm 0.090 $ & $ 46.28 \pm 0.19 $ \\
 6.512 & 2006 WO127 & 2009-Nov-04 01:09:24 & 16.18 & 1.528 & 0.889 & 40.23 & $ 0.102 \pm 0.011 $ & $ 0.350 \pm 0.018 $ \\
 6.516 & (140158) 2001 SX169 & 2010-Jan-26 04:30:24 & 18.30 & 1.253 & 0.430 & 47.17 & $ 0.092 \pm 0.010 $ & $ 0.362 \pm 0.018 $ \\
 6.526 & (100085) 1992 UY4 & 2010-Feb-07 01:04:57 & 17.80 & 1.226 & 0.832 & 54.76 & $ 0.317 \pm 0.017 $ & $ 1.694 \pm 0.037 $ \\
 6.534 & (85839) 1998 YO4 & 2010-Apr-05 10:55:52 & 16.30 & 1.326 & 0.783 & 48.87 & $ 0.179 \pm 0.014 $ & $ 0.792 \pm 0.026 $ \\
 6.539 & (68359) 2001 OZ13 & 2010-May-26 02:33:35 & 17.60 & 1.520 & 0.896 & 39.56 & $ 0.0243 \pm 0.0079 $ & $ 0.0443 \pm 0.0078 $ \\
 6.577 & (11398) 1998 YP11 & 2010-Apr-11 12:59:19 & 16.30 & 1.480 & 0.700 & 36.03 & $ 0.240 \pm 0.016 $ & $ 0.830 \pm 0.027 $ \\
 6.608 & (155334) 2006 DZ169 & 2009-Dec-20 23:59:59 & 17.10 & 1.636 & 0.871 & 32.65 & $ 0.0548 \pm 0.0072 $ & $ 0.148 \pm 0.011 $ \\
 6.609 & (175706) 1996 FG3 & 2010-May-02 05:19:51 & 18.20 & 1.213 & 0.418 & 51.02 & $ 0.853 \pm 0.028 $ & $ 4.391 \pm 0.062 $ \\
 6.610 & (52760) 1998 ML14 & 2010-Jan-26 05:06:11 & 17.50 & 1.075 & 0.484 & 69.33 & $ 0.127 \pm 0.012 $ & $ 0.463 \pm 0.020 $ \\
 6.628 & (138947) 2001 BA40 & 2010-Jan-10 17:16:01 & 18.40 & 1.173 & 0.721 & 59.14 & $ 0.0135 \pm 0.0035 $ & $ 0.0592 \pm 0.0071 $ \\
 6.652 & 2002 QE7 & 2009-Sep-23 04:42:21 & 19.33 & 1.217 & 0.329 & 47.87 & $ 0.0662 \pm 0.0080 $ & $ 0.188 \pm 0.013 $ \\
 6.652 & (5626) 1991 FE & 2009-Aug-21 03:23:36 & 14.70 & 1.689 & 0.972 & 33.08 & $ 0.434 \pm 0.020 $ & $ 1.209 \pm 0.031 $ \\
 6.661 & (66251) 1999 GJ2 & 2010-Feb-09 16:33:48 & 17.00 & 1.751 & 1.038 & 30.67 & $ 0.0294 \pm 0.0052 $ & $ 0.0512 \pm 0.0066 $ \\
 6.692 & (85990) 1999 JV6 & 2010-Jul-12 14:35:39 & 20.00 & 1.062 & 0.116 & 62.78 & $ 0.978 \pm 0.029 $ & $ 4.922 \pm 0.064 $ \\
 6.703 & 1998 SE36 & 2010-Apr-28 10:35:18 & 19.32 & 1.294 & 0.457 & 42.22 & $ 0.0280 \pm 0.0050 $ & $ 0.1089 \pm 0.0096 $ \\
 6.738 & (5645) 1990 SP & 2010-Jun-15 07:08:46 & 17.00 & 1.780 & 1.095 & 30.80 & $ 0.0496 \pm 0.0071 $ & $ 0.237 \pm 0.015 $ \\
 6.747 & (3671) Dionysus & 2010-Jun-12 23:01:07 & 16.30 & 1.136 & 0.518 & 62.60 & $ 0.180 \pm 0.013 $ & $ 0.390 \pm 0.019 $ \\
 6.750 & 2002 HF8 & 2009-Aug-24 23:44:37 & 18.27 & 1.262 & 0.449 & 48.69 & $ 0.1093 \pm 0.0099 $ & $ 0.455 \pm 0.020 $ \\
 6.751 & (164202) 2004 EW & 2009-Aug-06 05:32:19 & 20.80 & 1.049 & 0.158 & 75.37 & $ 0.0459 \pm 0.0062 $ & $ 0.135 \pm 0.010 $ \\
 6.757 & 2002 UN3 & 2010-Feb-12 22:31:31 & 18.60 & 1.325 & 0.503 & 41.88 & $ 0.0343 \pm 0.0058 $ & $ 0.0555 \pm 0.0070 $ \\
 6.791 & (90373) 2003 SZ219 & 2009-Dec-20 00:33:14 & 18.80 & 1.313 & 0.466 & 42.42 & $ 0.0332 \pm 0.0057 $ & $ 0.0686 \pm 0.0078 $ \\
 6.817 & (40329) 1999 ML & 2009-Sep-03 00:25:51 & 17.70 & 1.344 & 0.501 & 41.48 & $ 0.197 \pm 0.014 $ & $ 0.623 \pm 0.023 $ \\
 6.828 & (6239) Minos & 2010-Mar-20 17:55:44 & 17.90 & 1.122 & 0.340 & 61.27 & $ 0.114 \pm 0.012 $ & $ 0.28 \pm 0.17 $ \\
 6.830 & 2005 EJ & 2010-Mar-19 13:32:33 & 19.87 & 1.225 & 0.414 & 49.10 & $ 0.0215 \pm 0.0046 $ & $ 0.0530 \pm 0.0067 $ \\
 6.840 & (5646) 1990 TR & 2009-Dec-01 18:48:50 & 14.30 & 1.857 & 1.679 & 33.09 & $ 0.0393 \pm 0.0063 $ & $ 0.0809 \pm 0.0085 $ \\
 6.865 & 2003 WO7 & 2009-Oct-09 15:38:14 & 18.91 & 1.237 & 0.418 & 50.83 & $ 0.225 \pm 0.022 $ & $ 0.517 \pm 0.024 $ \\
 6.867 & 1999 RH33 & 2010-Apr-04 02:25:30 & 19.13 & 1.286 & 0.463 & 43.65 & $ 0.0177 \pm 0.0042 $ & $ 0.0371 \pm 0.0057 $ \\
 6.963 & (10115) 1992 SK & 2010-Apr-13 09:53:12 & 17.00 & 1.170 & 0.307 & 50.53 & $ 0.549 \pm 0.024 $ & $ 1.896 \pm 0.042 $ \\
 6.963 & 2003 BT47 & 2010-Apr-21 05:07:07 & 17.42 & 1.289 & 0.588 & 48.67 & $ 0.174 \pm 0.013 $ & $ 0.670 \pm 0.024 $ \\
 6.974 & (5620) Jasonwheeler & 2009-Aug-25 07:00:34 & 17.00 & 1.321 & 0.620 & 48.68 & $ 0.258 \pm 0.016 $ & $ 1.252 \pm 0.033 $ \\
 6.981 & (152563) 1992 BF & 2010-Apr-03 23:55:21 & 19.80 & 1.125 & 0.236 & 53.50 & $ 0.282 \pm 0.017 $ & $ 1.377 \pm 0.034 $ \\
 6.994 & (138971) 2001 CB21 & 2009-Dec-06 22:09:34 & 18.40 & 1.033 & 0.272 & 79.87 & $ 0.176 \pm 0.014 $ & $ 0.591 \pm 0.023 $ \\
\enddata

\tablecomments{ Times given are roughly mid-observation as measured on Spitzer. $H$ is the absolute optical magnitude (taken from \textit{Horizons} and assumed to be uncertain by $\pm0.5$~mag).  $r$ and $\Delta$ are heliocentric and Spitzer-centric distance, resp., $\alpha$ is the solar phase angle. f36 and f45 are the flux at $\sim3.6$ and \unit{4.5}{\micron}, resp.\ (channels 1 and 2).}
\end{deluxetable}

In this work, we  restrict ourselves to objects observed on or before 2010 July 14 with a rendezvous $\dv \leq \unit{7}{\km\per\second}$. \dv\ values are taken from Lance Benner's online list of \dv\ for all NEOs,%
\footnote{\url{http://echo.jpl.nasa.gov/~lance/delta_v/delta_v.rendezvous.html}}
which is calculated from the orbital elements and 
the  \citet{ShoemakerHelin1978} formalism. 

In Table \ref{tab:fluxes} we present the measured in-band fluxes and the observing circumstances as taken from JPL's \textit{Horizons} ephemeris server.  $H$ magnitudes are assumed to be uncertain by 0.5~mag 
(see Section \ref{sect:thermal}).
Observations carried out on or before 2009 November 4 have been presented in \citet{Trilling2010}, later observations are new here.

\section{Thermal modeling}
\label{sect:thermal}

We use an updated version of the  thermal-modeling pipeline used in 
\revision{previous publications resulting from}
the ExploreNEOs survey 
\citep{Trilling2010,Harris2011}.
For completeness, we briefly summarize the more detailed description given in  \citet{Trilling2010}.
Section \ref{sect:mc}  presents the updates relative to the previous pipeline, 
chiefly an estimation of the statistical uncertainty of our results.
Due to said update, some of our results differ slightly (but within the error bars) from the preliminary results given in \citet{Trilling2010}.  
A reanalysis of our entire data set is deferred to a later work.

NEO fluxes at Warm-Spitzer wavelengths have significant contributions from reflected sunlight and from thermal emission. We are interested in the latter in order to calculate the target size and albedo using a thermal model.  Therefore, in a first step, we estimate the amount of reflected solar radiation using the method first described by \citet{Mueller2007}, then refined in \citet{Trilling2008,Trilling2010}.
Briefly, we calculate the expected $V$ magnitude based on the observing circumstances given in Table \ref{tab:fluxes} and extrapolate to 3.6 and \unit{4.5}{\micron} fluxes using published values of the solar flux at those wavelengths and the Sun's $V$ magnitude. We also assume 
\revision{the spectral reflectivity at Warm Spitzer wavelengths to be 1.4 times that in the $V$ band}
\citep[see][]{Trilling2008,Harris2009,Trilling2010}.
In-band thermal flux equal\revision{s} total flux minus reflected flux.  In rare cases, the calculated reflected flux exceed\revision{s} the measured flux, leading to unphysical negative thermal fluxes in channel 1.  In these cases (which we expect are due to inaccurate $H$ magnitudes and/or lightcurve effects), we drop the channel-1 flux from the thermal analysis.

In-band thermal fluxes are color corrected to take account of 
\revision{the}
spectral breadth 
\revision{of IRAC's filters}
and the significant difference between the spectral shape of asteroidal thermal emission and the stellar-like spectrum assumed in IRAC flux calibration; color-correction factors for the reflected solar component are negligible.
Color-correction factors are calculated for each target using the  IRAC passbands given by \citet{Carey2010}, the observing circumstances, and a suitable thermal model.  As found by \citet{Mueller2007}, the dependence of color-correction factors on the physical properties of the asteroid can be neglected.

Diameter and albedo are estimated from the final thermal fluxes using the 
NEATM \citep{neatm}.
The NEATM contains a dimensionless parameter $\eta$ that describes the effective surface temperature. 
\citet{Trilling2008} found that 
\revision{data quality does not usually allow}
to fit $\eta$ to Warm-Spitzer data of NEOs, but that 
reasonable estimates of diameter and albedo can still be obtained by assuming
an empirical linear relationship between
$\eta$ and
solar phase angle $\alpha$.  
\revision{That}
relationship was 
established by \citet{Delbo2003}; here we  use an updated relationship (based on a slightly larger data set) by \citet{Wolters2008}: 
\begin{equation}
\label{eq:etaalpha}
\eta = (0.91\pm0.17) + (0.013\pm0.004) \alpha\ \textrm{(in deg.).}
\end{equation}


\subsection{Monte-Carlo approach}
\label{sect:mc}

\begin{figure}[!tb]
\includegraphics[angle=90,width=\linewidth]{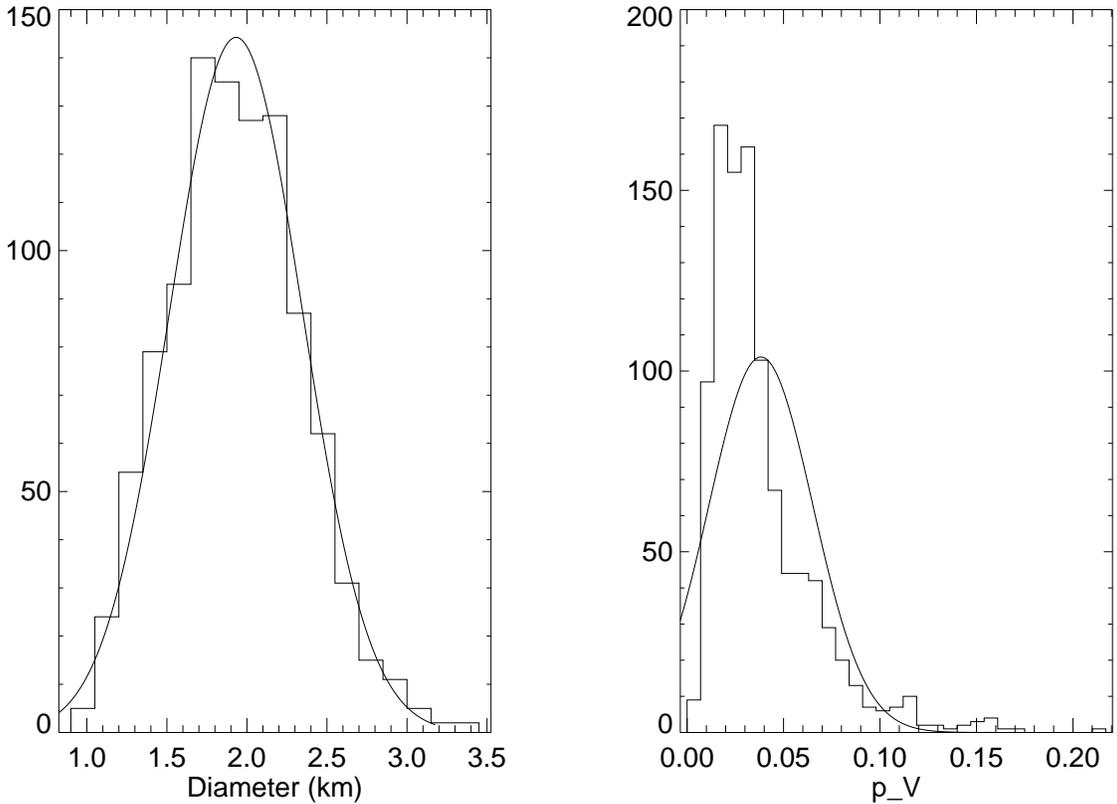}
\caption{Histogram of the distribution of diameter and albedo 
\revision{values}
 resulting from the Monte-Carlo procedure for 2001~SK162.  Both histograms are overplotted with the 
\revision{respective best-fit}
Gaussian functions.  Note that the diameter distribution is rather consistent with being normally distributed, while the albedo distribution is clearly not.
 \label{fig:histo}}
\end{figure}

In order to provide a realistic estimate of the uncertainty in our diameter and albedo results, we use a Monte-Carlo approach in which various sources of uncertainty are considered: the measured flux uncertainty, the calibration uncertainty of \unit{5}{\%} \citep{Carey2010}, the uncertainty in $H$ (see below), and the NEATM temperature parameter $\eta$ (which is assumed to vary by $\pm 0.3$ around its nominal value; see below).
For each observation, we generate 1000 sets of random synthetic fluxes normally distributed about the measured value and with a standard deviation equal to the 
\revision{root-sum-square}
of the measured flux uncertainty and the \unit{5}{\%} calibration uncertainty.
Analogously, Gaussian distributions of $H$ and $\eta$ values are used in the fit.  
%
The distribution of albedo results (diameters to a lesser extent) is strongly non-Gaussian, see Fig.\ \ref{fig:histo}.  We hence adopt the median of our Monte-Carlo results as the nominal value and asymmetric error bars to encompass the central \unit{68.2}{\%} of the results.
Additionally, we determine the percentage of albedo results falling into albedo bins (see Section \ref{sect:dpv}).

\paragraph{$\Delta H$:} 
In the analysis of purely thermal observations of asteroids, $H$ is known to have a negligible influence on best-fit diameters but impacts \pv\ directly \citep{harrisharris}.  In our case, however, $H$ is also used in correcting for reflected solar flux and hence influences the calculated thermal flux contribution.
We therefore decided to vary $H$  within the Monte-Carlo fit.
The correction for reflected sunlight 
\revision{becomes more critical as}
more reflected sunlight is contained within  the measured flux; it is hence
particularly important for high-albedo objects and for objects observed at large heliocentric distance.

Propagating $\Delta H$ into thermal-flux uncertainties is an update of the thermal-modeling pipeline relative to \citet{Trilling2010}.
While this does not change the calculated nominal thermal fluxes, it does change their uncertainties and hence the relative weight with which they enter the $\chi^2$ minimization procedure, 
 leading to somewhat different diameter and albedo results.
For practically all targets, however, the corresponding diameter change is $< \unit{1}{\%}$ and hence negligible.

For most of our targets, published values for $H$, let alone $\Delta H$, are unavailable.
An observing campaign to measure  $H$ for a number of our targets is currently underway; results will be reported in a separate paper.
For the time being, we fall back to the approximative $H$ values given by the \emph{Horizons} ephemeris server which
are notoriously unreliable%
\footnote{It should not be forgotten that \emph{Horizons} is not designed to calculate $H$ magnitudes but ephemerides, at which it does an excellent job.}
\citep[see][]{Juric2002,Parker2008}.
For now, we adopt $\Delta H = 0.5$~mag throughout, see Section \ref{sect:rerun} for exceptions.

\paragraph{$\Delta\eta$:}
Changes in the assumed $\eta$ value lead to significant changes in diameter and albedo, see e.g.\ \citet{Harris2011}.
The quoted uncertainties in Equation \eqref{eq:etaalpha} lead to a final $\Delta\eta\sim0.3$ for $\alpha\sim\unit{50}{\degree}$, a typical value for our sample.
This uncertainty estimate is corroborated by \citet{Ryan2010}, 
who found a typical $\eta$ value of $1.07\pm0.27$; we caution, however, that their sample is dominated by large main-belt asteroids, whose thermal properties are rather distinct from our sample of small NEOs.
For our Monte-Carlo analysis, we therefore adopt a Gaussian distribution of $\eta$ values scattering about the nominal value 
\revision{of $0.91+0.013\alpha$}
with a standard deviation of 0.3. Unphysical  $\eta$ values below 0.5 are discarded.

\section{Sizes and albedos}
\label{sect:dpv}

\revision{Our diameter and albedo results are given in Table~\ref{tab:results} along with their statistical uncertainties estimated from the Monte-Carlo analysis described above.}
\revision{
In order to illustrate the implications of our albedo results on surface mineralogy, we also determine the probabilities
$p_1$--$p_4$
with which the albedo falls within one of four albedo bins; these probabilities are estimated as the fraction of Monte-Carlo albedos falling within the respective bin.
The albedo bins are designed to correspond to taxonomic types as closely as possible, particularly for the purpose of identifying primitive NEOs:
}
\begin{itemize}
\item $\pv < 7.5\ \%$: As shown by \citet{Fernandez2005}, albedos in this range strongly indicate a primitive surface composition 
\item $7.5\ \% \leq \pv < 15\ \%$: While objects in this albedo range are still likely to be primitive, some of them may be more akin to (silicate rich) S types
\item $15\ \% \leq \pv < 30\ \%$ Objects in this albedo range are most likely S or Q types (or M types, but those are relatively rare among NEOs)
\item $30\ \% \leq \pv$ More exotic compositions, e.g.\ E types. 
\end{itemize}



\begin{deluxetable}{llllllllllllll}
\tabletypesize{\tiny}
\rotate
\tablecaption{\label{tab:results}Results of our Spitzer observations and of the dynamical analysis described in Section \ref{sect:history}
}
\tablehead{
\dv & Object & $D$ & \pv & $p_1$ & $p_2$ & $p_3$ & $p_4$ & $q_{10\%}$ & $q_{50\%}$ & $q_{90\%}$ & $T_{10\%}$ & $T_{50\%}$ & $T_{90\%}$ \\
(\kilo\metre\per\second) & & (\kilo\metre) & & & & & & (AU) & (AU) & (AU) & (\kelvin) & (\kelvin) & (\kelvin) 
}
\startdata
4.632 & (25143) Itokawa & $ 0.319^{+0.045}_{-0.050}$ & $ 0.41^{+0.20}_{-0.18}$ & 0.003 & 0.017 & 0.277 & 0.703 & 0.376 & 0.737 & 0.887 & 644 & 460 & 419 \\
4.755 & 1996 XB27 & $ 0.084^{+0.013}_{-0.012}$ & $ 0.48^{+0.26}_{-0.19}$ & 0.000 & 0.017 & 0.167 & 0.817 & 0.612 & 0.658 & 1.000 & 501 & 483 & 392 \\
4.887 & (10302) 1989 ML & $ 0.248^{+0.035}_{-0.043}$ & $ 0.47^{+0.28}_{-0.19}$ & 0.000 & 0.010 & 0.180 & 0.810 & 0.279 & 0.673 & 1.000 & 743 & 478 & 392 \\
5.276 & (99799) 2002 LJ3 & $ 0.503^{+0.094}_{-0.085}$ & $ 0.43^{+0.22}_{-0.18}$ & 0.000 & 0.033 & 0.233 & 0.733 & 0.313 & 0.665 & 0.875 & 704 & 483 & 421 \\
5.280 & 2001 CQ36 & $ 0.068^{+0.011}_{-0.012}$ & $ 0.41^{+0.29}_{-0.16}$ & 0.000 & 0.030 & 0.223 & 0.747 & 0.145 & 0.579 & 1.000 & 1038 & 519 & 395 \\
5.302 & (52381) 1993 HA & $ 0.337^{+0.097}_{-0.078}$ & $ 0.140^{+0.110}_{-0.077}$ & 0.210 & 0.337 & 0.357 & 0.097 & 0.258 & 0.668 & 0.874 & 802 & 498 & 435 \\
5.328 & 2000 YF29 & $ 0.244^{+0.041}_{-0.038}$ & $ 0.251^{+0.154}_{-0.095}$ & 0.003 & 0.140 & 0.487 & 0.370 & 0.229 & 0.663 & 0.838 & 840 & 494 & 439 \\
5.391 & (1943) Anteros & $ 2.48^{+0.69}_{-0.60}$ & $ 0.145^{+0.146}_{-0.073}$ & 0.183 & 0.330 & 0.350 & 0.137 & 0.313 & 0.665 & 0.875 & 727 & 499 & 435 \\
5.486 & (138911) 2001 AE2 & $ 0.352^{+0.073}_{-0.069}$ & $ 0.34^{+0.22}_{-0.16}$ & 0.000 & 0.090 & 0.337 & 0.573 & 0.310 & 0.627 & 0.885 & 715 & 503 & 423 \\
5.487 & 2006 SY5 & $ 0.090^{+0.013}_{-0.017}$ & $ 0.34^{+0.23}_{-0.14}$ & 0.007 & 0.083 & 0.320 & 0.590 & 0.302 & 0.572 & 0.879 & 725 & 527 & 425 \\
5.555 & (162416) 2000 EH26 & $ 0.141^{+0.026}_{-0.024}$ & $ 0.181^{+0.142}_{-0.073}$ & 0.053 & 0.313 & 0.443 & 0.190 & 0.289 & 0.775 & 0.901 & 754 & 460 & 427 \\
5.565 & (162998) 2001 SK162 & $ 1.94^{+0.38}_{-0.37}$ & $ 0.031^{+0.028}_{-0.015}$ & 0.920 & 0.077 & 0.003 & 0.000 & 0.244 & 0.793 & 0.967 & 834 & 462 & 418 \\
5.653 & (68372) 2001 PM9 & $ 1.73^{+0.45}_{-0.41}$ & $ 0.0180^{+0.0170}_{-0.0080}$ & 0.970 & 0.030 & 0.000 & 0.000 & 0.278 & 0.711 & 0.872 & 782 & 488 & 441 \\
5.719 & (12923) Zephyr & $ 1.86^{+0.45}_{-0.46}$ & $ 0.21^{+0.17}_{-0.11}$ & 0.070 & 0.247 & 0.410 & 0.273 & 0.304 & 0.714 & 0.903 & 733 & 478 & 425 \\
6.041 & (85938) 1999 DJ4 & $ 0.478^{+0.100}_{-0.082}$ & $ 0.28^{+0.23}_{-0.13}$ & 0.017 & 0.140 & 0.393 & 0.450 & 0.303 & 0.690 & 0.882 & 729 & 482 & 427 \\
6.069 & (433) Eros & $ 30.7^{+10.0}_{-8.3}$ & $ 0.065^{+0.075}_{-0.034}$ & 0.560 & 0.303 & 0.117 & 0.020 & 0.320 & 0.580 & 0.742 & 724 & 538 & 476 \\
6.070 & 2000 GV147 & $ 0.502^{+0.104}_{-0.072}$ & $ 0.185^{+0.124}_{-0.084}$ & 0.063 & 0.300 & 0.460 & 0.177 & 0.263 & 0.577 & 0.820 & 791 & 534 & 447 \\
6.086 & (217807) 2000 XK44 & $ 0.73^{+0.14}_{-0.13}$ & $ 0.28^{+0.18}_{-0.12}$ & 0.013 & 0.130 & 0.387 & 0.470 & 0.319 & 0.617 & 0.925 & 711 & 510 & 417 \\
6.106 & 1993 RA & $ 0.358^{+0.063}_{-0.052}$ & $ 0.40^{+0.23}_{-0.16}$ & 0.007 & 0.040 & 0.300 & 0.653 & 0.323 & 0.713 & 0.924 & 697 & 468 & 411 \\
6.130 & (177614) 2004 HK33 & $ 0.94^{+0.17}_{-0.18}$ & $ 0.189^{+0.141}_{-0.081}$ & 0.073 & 0.267 & 0.457 & 0.203 & 0.306 & 0.638 & 0.820 & 732 & 507 & 447 \\
6.191 & (192559) 1998 VO & $ 0.216^{+0.032}_{-0.032}$ & $ 0.30^{+0.17}_{-0.13}$ & 0.007 & 0.107 & 0.380 & 0.507 & 0.247 & 0.521 & 0.745 & 805 & 554 & 463 \\
6.196 & (212546) 2006 SV19 & $ 1.06^{+0.23}_{-0.21}$ & $ 0.129^{+0.104}_{-0.058}$ & 0.183 & 0.420 & 0.297 & 0.100 & 0.216 & 0.809 & 1.000 & 876 & 453 & 407 \\
6.236 & 2005 JA22 & $ 0.67^{+0.17}_{-0.14}$ & $ 0.164^{+0.117}_{-0.078}$ & 0.120 & 0.320 & 0.423 & 0.137 & 0.200 & 0.747 & 0.970 & 908 & 470 & 412 \\
6.240 & (1627) Ivar & $ 9.4^{+3.9}_{-2.3}$ & $ 0.094^{+0.138}_{-0.051}$ & 0.390 & 0.297 & 0.253 & 0.060 & 0.347 & 0.681 & 0.840 & 695 & 496 & 446 \\
6.279 & (87024) 2000 JS66 & $ 0.312^{+0.059}_{-0.039}$ & $ 0.63^{+0.34}_{-0.23}$ & 0.000 & 0.003 & 0.057 & 0.940 & 0.279 & 0.598 & 0.874 & 728 & 497 & 411 \\
6.323 & (22099) 2000 EX106 & $ 0.621^{+0.109}_{-0.076}$ & $ 0.29^{+0.16}_{-0.12}$ & 0.003 & 0.113 & 0.413 & 0.470 & 0.138 & 0.530 & 0.804 & 1080 & 550 & 446 \\
6.364 & 2003 SL5 & $ 0.337^{+0.062}_{-0.053}$ & $ 0.38^{+0.22}_{-0.16}$ & 0.003 & 0.027 & 0.350 & 0.620 & 0.269 & 1.000 & 1.000 & 764 & 396 & 396 \\
6.364 & (35107) 1991 VH & $ 1.12^{+0.23}_{-0.20}$ & $ 0.27^{+0.20}_{-0.12}$ & 0.023 & 0.143 & 0.413 & 0.420 & 0.094 & 0.525 & 0.884 & 1311 & 554 & 427 \\
6.379 & (172974) 2005 YW55 & $ 0.342^{+0.079}_{-0.059}$ & $ 0.30^{+0.24}_{-0.13}$ & 0.000 & 0.103 & 0.383 & 0.513 & 0.269 & 0.901 & 1.000 & 772 & 421 & 400 \\
6.405 & (65679) 1989 UQ & $ 0.73^{+0.18}_{-0.15}$ & $ 0.060^{+0.059}_{-0.028}$ & 0.647 & 0.270 & 0.077 & 0.007 & 0.144 & 0.530 & 1.000 & 1082 & 564 & 410 \\
6.431 & (159402) 1999 AP10 & $ 1.20^{+0.29}_{-0.17}$ & $ 0.35^{+0.23}_{-0.16}$ & 0.007 & 0.057 & 0.337 & 0.600 & 0.148 & 0.728 & 0.933 & 1033 & 466 & 412 \\
6.491 & (143651) 2003 QO104 & $ 2.29^{+0.54}_{-0.51}$ & $ 0.137^{+0.140}_{-0.061}$ & 0.160 & 0.380 & 0.327 & 0.133 & 0.243 & 0.639 & 1.000 & 826 & 509 & 407 \\
6.507 & 1989 AZ & $ 1.09^{+0.20}_{-0.19}$ & $ 0.025^{+0.019}_{-0.011}$ & 0.967 & 0.030 & 0.003 & 0.000 & 0.229 & 0.541 & 0.791 & 861 & 560 & 463 \\
6.512 & (218863) 2006 WO127 & $ 1.70^{+0.40}_{-0.34}$ & $ 0.21^{+0.17}_{-0.11}$ & 0.060 & 0.227 & 0.443 & 0.270 & 0.213 & 0.622 & 0.924 & 875 & 512 & 420 \\
6.516 & (140158) 2001 SX169 & $ 0.58^{+0.16}_{-0.12}$ & $ 0.26^{+0.19}_{-0.11}$ & 0.033 & 0.150 & 0.433 & 0.383 & 0.204 & 0.454 & 0.667 & 891 & 596 & 492 \\
6.526 & (100085) 1992 UY4 & $ 2.60^{+0.70}_{-0.64}$ & $ 0.020^{+0.022}_{-0.010}$ & 0.943 & 0.050 & 0.007 & 0.000 & 0.684 & 0.862 & 0.977 & 498 & 444 & 417 \\
6.534 & (85839) 1998 YO4 & $ 1.81^{+0.42}_{-0.36}$ & $ 0.174^{+0.136}_{-0.086}$ & 0.103 & 0.310 & 0.417 & 0.170 & 0.269 & 0.901 & 1.000 & 783 & 427 & 406 \\
\tablebreak
6.539 & (68359) 2001 OZ13 & $ 0.62^{+0.15}_{-0.12}$ & $ 0.42^{+0.27}_{-0.19}$ & 0.007 & 0.047 & 0.240 & 0.707 & 0.209 & 0.625 & 0.873 & 864 & 499 & 422 \\
6.577 & (11398) 1998 YP11 & $ 1.73^{+0.47}_{-0.40}$ & $ 0.176^{+0.185}_{-0.090}$ & 0.097 & 0.317 & 0.343 & 0.243 & 0.400 & 0.754 & 1.000 & 641 & 467 & 405 \\
6.608 & (155334) 2006 DZ169 & $ 1.15^{+0.32}_{-0.29}$ & $ 0.195^{+0.183}_{-0.095}$ & 0.080 & 0.253 & 0.387 & 0.280 & 0.275 & 0.803 & 1.000 & 771 & 451 & 404 \\
6.609 & (175706) 1996 FG3 & $ 1.90^{+0.52}_{-0.44}$ & $ 0.026^{+0.029}_{-0.012}$ & 0.923 & 0.067 & 0.010 & 0.000 & 0.307 & 0.531 & 0.730 & 743 & 565 & 482 \\
6.610 & (52760) 1998 ML14 & $ 0.81^{+0.16}_{-0.14}$ & $ 0.27^{+0.24}_{-0.11}$ & 0.007 & 0.140 & 0.413 & 0.440 & 0.456 & 0.731 & 0.895 & 594 & 469 & 424 \\
6.628 & (138947) 2001 BA40 & $ 0.440^{+0.085}_{-0.080}$ & $ 0.42^{+0.26}_{-0.18}$ & 0.003 & 0.053 & 0.213 & 0.730 & 0.205 & 0.484 & 0.723 & 871 & 567 & 464 \\
6.652 & 2002 QE7 & $ 0.320^{+0.061}_{-0.056}$ & $ 0.34^{+0.20}_{-0.14}$ & 0.010 & 0.073 & 0.337 & 0.580 & 0.318 & 0.598 & 0.829 & 707 & 515 & 437 \\
6.652 & (5626) 1991 FE & $ 3.96^{+1.22}_{-0.92}$ & $ 0.152^{+0.159}_{-0.081}$ & 0.173 & 0.323 & 0.340 & 0.163 & 0.644 & 1.000 & 1.000 & 507 & 406 & 406 \\
6.661 & (66251) 1999 GJ2 & $ 0.90^{+0.24}_{-0.19}$ & $ 0.37^{+0.30}_{-0.19}$ & 0.013 & 0.067 & 0.307 & 0.613 & 0.065 & 0.556 & 0.878 & 1558 & 533 & 424 \\
6.692 & (85990) 1999 JV6 & $ 0.498^{+0.134}_{-0.088}$ & $ 0.076^{+0.058}_{-0.035}$ & 0.500 & 0.373 & 0.120 & 0.007 & 0.142 & 0.457 & 0.629 & 1086 & 606 & 516 \\
6.703 & 1998 SE36 & $ 0.343^{+0.069}_{-0.072}$ & $ 0.30^{+0.22}_{-0.14}$ & 0.027 & 0.127 & 0.347 & 0.500 & 0.209 & 0.348 & 0.666 & 875 & 678 & 490 \\
6.738 & (5645) 1990 SP & $ 2.20^{+0.74}_{-0.64}$ & $ 0.062^{+0.079}_{-0.034}$ & 0.597 & 0.260 & 0.110 & 0.033 & 0.305 & 0.576 & 0.811 & 743 & 540 & 456 \\
6.747 & (3671) Dionysus & $ 0.89^{+0.11}_{-0.11}$ & $ 0.67^{+0.37}_{-0.24}$ & 0.000 & 0.000 & 0.060 & 0.940 & 0.243 & 0.639 & 1.000 & 776 & 479 & 382 \\
6.750 & 2002 HF8 & $ 0.71^{+0.16}_{-0.15}$ & $ 0.181^{+0.138}_{-0.081}$ & 0.090 & 0.300 & 0.410 & 0.200 & 0.759 & 1.000 & 1.000 & 465 & 405 & 405 \\
6.751 & (164202) 2004 EW & $ 0.157^{+0.024}_{-0.021}$ & $ 0.36^{+0.22}_{-0.15}$ & 0.000 & 0.043 & 0.317 & 0.640 & 0.144 & 0.530 & 1.000 & 1048 & 546 & 397 \\
6.757 & 2002 UN3 & $ 0.310^{+0.052}_{-0.047}$ & $ 0.67^{+0.35}_{-0.26}$ & 0.000 & 0.000 & 0.070 & 0.930 & 0.609 & 0.912 & 1.000 & 490 & 400 & 382 \\
6.791 & (90373) 2003 SZ219 & $ 0.306^{+0.049}_{-0.049}$ & $ 0.58^{+0.30}_{-0.23}$ & 0.000 & 0.013 & 0.060 & 0.927 & 0.269 & 0.901 & 1.000 & 747 & 408 & 387 \\
6.817 & (40329) 1999 ML & $ 0.96^{+0.23}_{-0.23}$ & $ 0.154^{+0.157}_{-0.069}$ & 0.117 & 0.370 & 0.347 & 0.167 & 0.608 & 0.994 & 1.000 & 521 & 408 & 406 \\
6.828 & (6239) Minos & $ 0.474^{+0.117}_{-0.091}$ & $ 0.56^{+0.39}_{-0.27}$ & 0.000 & 0.040 & 0.123 & 0.837 & 0.301 & 0.487 & 0.694 & 708 & 556 & 466 \\
6.830 & 2005 EJ & $ 0.230^{+0.041}_{-0.038}$ & $ 0.43^{+0.24}_{-0.19}$ & 0.000 & 0.033 & 0.243 & 0.723 & 0.318 & 0.598 & 0.829 & 700 & 510 & 433 \\
6.840 & (5646) 1990 TR & $ 2.03^{+0.52}_{-0.28}$ & $ 0.65^{+0.43}_{-0.28}$ & 0.000 & 0.020 & 0.080 & 0.900 & 0.400 & 1.000 & 1.000 & 606 & 383 & 383 \\
6.865 & 2003 WO7 & $ 0.68^{+0.16}_{-0.13}$ & $ 0.109^{+0.074}_{-0.051}$ & 0.277 & 0.480 & 0.210 & 0.033 & 0.400 & 1.000 & 1.000 & 646 & 408 & 408 \\
6.867 & 1999 RH33 & $ 0.228^{+0.042}_{-0.032}$ & $ 0.73^{+0.35}_{-0.27}$ & 0.000 & 0.000 & 0.043 & 0.957 & 0.065 & 0.556 & 0.878 & 1488 & 509 & 405 \\
6.963 & 2003 BT47 & $ 1.15^{+0.31}_{-0.24}$ & $ 0.147^{+0.149}_{-0.074}$ & 0.173 & 0.340 & 0.340 & 0.147 & 0.650 & 0.901 & 1.000 & 504 & 428 & 406 \\
6.963 & (10115) 1992 SK & $ 0.90^{+0.20}_{-0.18}$ & $ 0.34^{+0.25}_{-0.13}$ & 0.003 & 0.050 & 0.360 & 0.587 & 0.129 & 0.491 & 0.754 & 1109 & 568 & 459 \\
6.974 & (5620) Jasonwheeler & $ 1.77^{+0.46}_{-0.40}$ & $ 0.094^{+0.096}_{-0.046}$ & 0.360 & 0.373 & 0.223 & 0.043 & 0.400 & 1.000 & 1.000 & 647 & 409 & 409 \\
6.981 & (152563) 1992 BF & $ 0.51^{+0.12}_{-0.11}$ & $ 0.084^{+0.077}_{-0.037}$ & 0.410 & 0.407 & 0.147 & 0.037 & 0.242 & 0.514 & 0.686 & 833 & 571 & 494 \\
6.994 & (138971) 2001 CB21 & $ 0.578^{+0.109}_{-0.079}$ & $ 0.24^{+0.12}_{-0.10}$ & 0.020 & 0.213 & 0.487 & 0.280 & 0.142 & 0.457 & 0.629 & 1068 & 596 & 508 \\
\enddata
 
\tablecomments{
$p_1$--$p_4$ denote the probability of the albedo falling within each of the four albedo bins described in the text (bin boundaries are 0.075, 0.15, and 0.3; primitive objects should display a high $p_1$).
The last six columns, $q_{xx}$ and $T_{xx}$ describe the orbital and thermal history (see Section \ref{sect:history}). E.g., $q_{10\%}$ denotes the minimum perihelion distance that the object reached to within a probability of 10\%, $T_{10\%}$ is the corresponding temperature to which the surface was heated.
Note that some of the diameter and albedo results herein are superseded  in Table \ref{tab:newh}.
}
\clearpage
\end{deluxetable}

\subsection{Reanalysis with updated $H$ and $G$ values for select objects}
\label{sect:rerun}

\begin{deluxetable}{llllllll}
\tabletypesize{\footnotesize}
\tablecaption{
\label{tab:hgd} %
Published physical properties 
for objects considered in Sect.\ \ref{sect:rerun}}
\tablehead{Object & $H$ & $G$ & $D$(\kilo\metre) & $\pv$ & Bin\tablenotemark{a} & Taxo\tablenotemark{b} & Reference}
\startdata
(433) Eros\tablenotemark{c} & $10.46\pm0.10$ & 0.18 & 23.6\tablenotemark{a} & 0.22\tablenotemark{b} &-- & S & 1,2\\
(25143) Itokawa & $19.51^{+0.09}_{-0.08}$\tablenotemark{c} & $0.29^{+0.07}_{-0.06}$& $0.327\pm0.006$ & $0.26\pm0.02$ & --& S(IV) & 3,4\\
(10302) 1989 ML & --- &--- & $0.28\pm0.05$& $0.37\pm0.15$&-- & E& 5\\
(175706) 1996 FG3 & $17.76\pm0.03$& $-0.07$ & $\sim1.9$& $\sim0.04$& Y& C& 6,7,8\\
(1943) Anteros & $15.9\pm0.2$ & 0.23 & $\sim2.2$ & $\sim0.16$ & --& L& 9,10,11\\
(65679) 1989 UQ & $19.5\pm0.3$ & ---& ---& ---& --& B & 11,12\\
(162998) 2001 SK162 & --- & --- & --- & --- & --& T & 11\\
(100085) 1992 UY4   & $17.71\pm0.10$ &--- & 1.7 & 0.05 & -- & P & 13,14\\
(152563) 1992 BF & ---& ---& ---& ---& -- & $X_c$ & 15\\
(12923) Zephyr   &---&---&---&---&--& S& 16 \\
(1627) Ivar      & $12.87\pm0.10$  & --- & $9.1\pm1.4$ & $0.15\pm0.05$ &  -- &S & 16,17\\
(85990) 1999 JV6 & --- & --- & --- & --- & -- & $X_k$ & 15\\
(3671) Dionysus  & $16.66\pm0.30$ & --- & 1.1--1.5 & 0.16--0.31 & Y & $C_b$ & 15,18,19\\
\enddata
\tablenotetext{a}{Y indicates that the object is known to be binary}
\tablenotetext{b}{Taxonomic type}
\tablenotetext{c}{See \citet{Trilling2010} for a discussion of the values adopted for Eros, which was observed at a nearly pole-on viewing geometry, and \citet{MuellerThesis} for a discussion of its $G$ value.}
\tablerefs{
(1) \citet{harrisdavies};
(2) \citet{li04};
(3) \citet{Bernardi2008};
(4) Volume-equivalent diameter calculated from the  volume given by \citet{Demura2006}; 
(5) \citet{Mueller2007};
(6) \citet{Pravec2006};
(7) Mueller et al., in preparation;
(8) \citet{Binzel2001};
(9) Adopted after $15.82\pm0.14$\citep{Wisniewski1997} and $15.96\pm0.14$ \citep{Pravec1998}, both give $G=0.23$;
(10) Quoted after \citet{Harris2011}, original data from \citet{Veeder1989};
(11) \citet{Binzel2004};
(12) \citet{Pravec1998};
(13) \citet{Warner2006}---no $H$ uncertainty is given by \citeauthor{Warner2006}, 0.1~mag seems appropriate (or slightly conservative) given the low scatter of their data;
(14) \citet{Volquardsen2007}---note that their error bars (not quoted herein) reflect only the statistical uncertainties, but not the systematics;
(15) \citet{BusBinzel2002};
(16) \citet{Binzel2004b};
(17) \citet{Delbo2003} determined $D$ and \pv\ of Ivar based on Keck mid-IR photometry, we here assume a 15\% uncertainty in $D$ and 30\% in \pv\ as is usual for radiometric diameters.  $H$ is from Pravec et al.\ (unpublished), quoted after \citet{Delbo2003}.
(18) Different, model-dependent, diameter and albedo values are given by \citet{harrisdavies,Harris2002}---we quote the range of their adopted results.
\citet{harrisdavies} quote Pravec (unpublished) for the $H$ given herein, no uncertainty value is stated, we assume a conservative uncertainty of 0.3~mag.
(19) \citet{Mottola1997}
}
\end{deluxetable}


In the analysis above, we assume $G=0.15$ and use $H$ magnitudes from the \emph{Horizons} ephemeris service.
While both assumptions are known to be problematic, we are constrained to use them in the 'mass production' of diameters and albedo for practical reasons: 
\revision{w}e
are not aware of a central database of published $H$ and $G$ values; rather, values for each target have to be searched in the literature and are unavailable for a large majority.
In order to minimize  the induced uncertainties, we reanalyze our data
for objects with published  $H$ and/or $G$ values; see Table \ref{tab:hgd}.  We focus on objects which we 
\revision{find}
 to have low albedo, as well as Eros and Itokawa. Where available, we also include 
published determinations of the size, albedo, or taxonomic type.
The results of this reanalysis are given in Table \ref{tab:newh}.  

\begin{deluxetable}{lllllllll}
\tabletypesize{\footnotesize}
\tablecaption{
\label{tab:newh}%
Reanalysis of thermal data with updated $H$ and $G$ values (see Table \ref{tab:hgd})
}
\tablehead{
Object & $D$(\kilo\metre) & \pv & $p_1$ & $p_2$ & $p_3$ & $p_4$ & $D^*$ & $\pv^*$
}
\startdata
(433) Eros & $ 23.0^{+8.3}_{-5.6}$ & $ 0.218^{+0.173}_{-0.099}$ & 0.030 & 0.250 & 0.423 & 0.297 & $ 30.7^{+10.0}_{-8.3}$ & $ 0.065^{+0.075}_{-0.034}$ \\
(25143) Itokawa & $ 0.313^{+0.054}_{-0.044}$ & $ 0.283^{+0.116}_{-0.075}$ & 0.000 & 0.023 & 0.550 & 0.427 & $ 0.319^{+0.045}_{-0.050}$ & $ 0.41^{+0.20}_{-0.18}$ \\
(10302) 1989 ML & $ 0.240^{+0.043}_{-0.038}$ & $ 0.50^{+0.21}_{-0.18}$ & 0.003 & 0.010 & 0.127 & 0.860 & $ 0.248^{+0.035}_{-0.043}$ & $ 0.47^{+0.28}_{-0.19}$ \\
(175706) 1996 FG3 & $ 1.84^{+0.56}_{-0.47}$ & $ 0.042^{+0.035}_{-0.017}$ & 0.837 & 0.140 & 0.023 & 0.000 & $ 1.90^{+0.52}_{-0.44}$ & $ 0.026^{+0.029}_{-0.012}$ \\
(1943) Anteros & $ 2.38^{+0.72}_{-0.59}$ & $ 0.138^{+0.107}_{-0.061}$ & 0.150 & 0.420 & 0.367 & 0.063 & $ 2.48^{+0.69}_{-0.60}$ & $ 0.145^{+0.146}_{-0.073}$ \\
(65679) 1989 UQ & $ 0.72^{+0.18}_{-0.14}$ & $ 0.053^{+0.036}_{-0.021}$ & 0.753 & 0.223 & 0.023 & 0.000 & $ 0.73^{+0.18}_{-0.15}$ & $ 0.060^{+0.059}_{-0.028}$ \\
(100085) 1992 UY4 & $ 2.50^{+0.67}_{-0.58}$ & $ 0.0230^{+0.0190}_{-0.0090}$ & 0.977 & 0.020 & 0.003 & 0.000 & $ 2.60^{+0.70}_{-0.64}$ & $ 0.020^{+0.022}_{-0.010}$ \\
(1627) Ivar & $ 9.9^{+2.8}_{-2.8}$ & $ 0.128^{+0.123}_{-0.052}$ & 0.157 & 0.447 & 0.273 & 0.123 & $ 9.4^{+3.9}_{-2.3}$ & $ 0.094^{+0.138}_{-0.051}$ \\
(3671) Dionysus & $ 0.86^{+0.12}_{-0.11}$ & $ 0.55^{+0.21}_{-0.17}$ & 0.000 & 0.000 & 0.047 & 0.953 & $ 0.89^{+0.11}_{-0.11}$ & $ 0.67^{+0.37}_{-0.24}$ \\
Dionysus with $\eta=3$ & $ 1.46^{+0.21}_{-0.18}$ & $ 0.179^{+0.092}_{-0.065}$ & 0.020 & 0.320 & 0.553 & 0.107 &\nodata&\nodata\\
\enddata
 

\tablecomments{
In the last two columns, the diameter and albedo results for the 'nominal' $H$ and $G$ values are given for comparison (see Table \ref{tab:results}).
\revision{In this reanalysis, $\eta=1.07$ is assumed for (433) Eros}
\citep[see][]{Trilling2010}.
Two $\eta$ assumptions are made for (3671) Dionysus, the one used in the rest of this manuscript (see Equation \eqref{eq:etaalpha}; upper line), and $\eta=3$ (lower line).
\revision{
For all objects except Eros and Dionysus, 
$D$ and $D^*$ values are practically indistinguishable, implying that diameter results are not significantly impacted by the $H$ uncertainty.}
}
\end{deluxetable}

\paragraph{Low-albedo objects:}

For (175706) 1996~FG3, (65679) 1989~UQ, and (100085) 1992~UY4, our albedo results are low, as expected based on their known taxonomic classification.  In the case of 1996~FG3, there is also an excellent quantitative match between our results and the ground-based measurements quoted in Table \ref{tab:hgd}.  Our diameter result for 1992~UY4 is  formally in agreement with that by \citet{Volquardsen2007}, provided that 
\revision{their}
value is assigned a realistic uncertainty of $\sim\unit{15}{\%}$ to include systematic uncertainties (which 
\revision{are}
not discussed by the authors of that paper).

\paragraph{'Reality checks:'}

While the general agreement between ExploreNEOs results and other published diameter and albedo results is discussed
 by \citet{Harris2011}, we consider it useful to check some of our results for low-\dv\ objects against published values.
By comparing Tables \ref{tab:hgd} and \ref{tab:newh}, we find excellent agreement in the cases of (433) Eros, (25143) Itokawa, (10302) 1989 ML, (1943) Anteros, and (1627) Ivar.


\paragraph{(3671) Dionysus:}

\revision{Our nominal albedo result for Dionysus, $\pv=0.55^{+0.21}_{-0.17}$, is hard to reconcile with its taxonimic classification as $C_b$ type, for which a low albedo  would be expected.
\citet{harrisdavies} report UKIRT mid-IR observations of Dionysus from which they derive $\pv=0.35$ or $\pv=0.61$, depending on thermal model. Due to ``inadequate signal-to-noise'', they were unable to constrain $\eta$ from their data, like in our case.
However, \citeauthor{harrisdavies} also report ISO data, from which $\eta$ can be constrained (if marginally so) to be $\sim 3.1$,
in the upper range of plausible $\eta$ values.
This results in
 $\pv=0.16$ \citep[quoted after][---\citeauthor{harrisdavies} reject that result in favor of another thermal model, leading to much higher albedo]{Harris2002}. 
}
\revision{We have therefore repeated}
the analysis of our Dionysus data assuming $\eta=3$ (
the phase angle of our observations,  $\alpha=\unit{62.6}{\degree}$, is very similar to that of 
\revision{the ISO observations of}
\citeauthor{harrisdavies}, \unit{57.7}{\degree}).
\revision{The resulting albedo, $\pv=0.178^{+0.092}_{-0.065}$, is in good agreement with Harris' NEATM result ($\pv=0.16$)}
and consistent with a $C_b$ classification given the error bars.

We note that binary NEOs including Dionysus were recently found to generally display higher-than-average $\eta$ values \citep{Delbo2011}, probably due to regolith loss during binary formation.
This may be expected to reduce the accuracy of our results for binaries in general.  
However, in the case of the only other known low-albedo binary, 1996~FG3 (see above), our results are quite consistent with those obtained otherwise.

\section{Thermal history}
\label{sect:history}
It is well known that implantation of 
\revision{solar wind ions}
\citep{Hapke:2001p5922} and 
bombardment 
\revision{by}
micrometeorites can
alter the spectroscopic properties of asteroids
\citep{Sasaki:3p5923}. However, these aging processes affect only the
\revision{topmost}
 microns of the surface. This is not a problem for a sample
collection experiment: current-technology sampling devices
can sample material from a depth of a few centimeters, thus excavating 
below the space-weathered surface. 

However, \citet{Marchi:2009p5116} have shown that the surfaces of a
significant number of NEOs 
\revision{were}
heated by the Sun to 
\revision{very high}
temperatures 
\revision{that could}
induce surface alterations on previously primitive
objects.  
\revision{
Due to thermal conduction, a thin layer beneath the surface can be heated to similarly large temperatures; for}
typical thermal properties
\citep{MuellerThesis,Delbo2007} and rotation periods, the penetration
depth of the heat wave is of the order of centimeters
\citep{Spencer:1989p482}, comparable to the digging depth of
sample-taking devices.

There is no clear-cut way to determine the maximum 
temperature to which an asteroid can be heated and remain in 
primitive condition.
From laboratory studies of carbonaceous chondrite meteorites, 
to which primitive asteroids are generally believed to be related, 
we know different alteration processes that are characterized by 
different threshold temperatures (e.g. thermal breakup of 
organic macromolecules).
%
\revision{ For instance, it has been determined by laboratory heating
  experiments that at \unit{370}{\kelvin} the insoluble organic matter of the
  carbonaceous meteorite Murchinson is degraded (the aliphatic C-H
  bond is lost) in approximately 200 years \citep{Kebukawa2010}. The
  same authors also showed that the bulk organics of Murchinson are
  lost in only one year at 370 K (or 200 years at 300 K).}  It is also
known that the macromolecular phase in carbonaceous meteorites has a
structure similar to refractory kerogen. The latter starts to break up
- with production of oil and gas - when heated above
\unit{420}{\kelvin} 
\revision{(I.\ Franchi 2008, personal communication)}.  \revision{Furthermore, \cite{Lauretta2001} have
  shown experimentally that volatile components (such as Hg) are
  released from some CM and CV carbonaceous chondrites when the latter are heated
above $\sim 470$ K.}

The maximum temperature attained by an NEO is a function of the
perihelion distance
$q$. 
NEO orbits can evolve rapidly \citep[see, e.g.,][]{Michel:1997p5919},
hence the present $q$ is not particularly indicative of the minimum
$q$ attained within the chaotic dynamical history.
\revision{Hence, if}
it is a goal to send a spacecraft to a primitive object, the dynamical 
and thermal history of the target must be taken into account.


An upper limit on the subsurface temperature is the surface temperature at local noon.
Because  spin axes of NEOs evolve on relatively short timescales  (e.g.\ due to
YORP, planetary encounters, and possibly due to spin-orbit coupling)
the whole surface is likely to have been subjected to temperatures
(nearly) as high as that of the subsolar point, $T_{SS}$.
We calculate $T_{SS}$ as a function of $q$ using
the NEATM (see Section \ref{sect:thermal}), assuming the nominal albedo resulting from our Spitzer observations and $\eta=0.91$. 
That latter value follows from Equation \eqref{eq:etaalpha} for $\alpha=0$, hence providing an upper limit on temperature.
%

With this in mind, the maximum surface temperature attained follows 
from the minimum $q$ reached in the past.  Due to the chaotic 
nature of NEO orbits, this question must be treated probabilistically.
We here use the orbital evolution model by
\citet{Marchi:2009p5116}, which was derived from
\citet{BottkeJr:2000p5926,Bottke:2002p5583}. 
For each of our targets, we extract from the work by 
\citeauthor{Marchi:2009p5116}\ the probabilities that the 
object reached a perihelion distance smaller than a grid of $q$ values.
Through interpolation, we determine $q_{50\%}$, i.e.\
 the $q$ value 
which was reached with a probability of 50\%.
We call the corresponding subsolar temperature $T_{50\%}$. 
The odds are 50\% that an object ventured to within 
$q_{50\%}$ of the Sun and that its surface was heated to $T_{50\%}$ or more.
We repeat this exercise for probability values of 10 and 90\%; 
results are given in Table \ref{tab:results}.

Note, for instance, that (65679) 1989 UQ  certainly appears primitive
judging from its  low albedo (this work) and its
spectroscopic classification 
\revision{as B type} 
\citep{Binzel2004}. 
1989~UQ was considered repeatedly as a target of sample-return
 mission concepts to a primitive asteroid.  However,  we  show
that this object has likely been heated to the point that its
organic macromolecular matter has been broken up 
($T_{50\%} = \unit{564}{\kelvin}$---cf.\ Table \ref{tab:results}).
The same applies to the binary system (175706) 1996 FG3.

However, the final determination whether an object is to be considered primitive or not is beyond our scope. Rather, our aim is to provide the required input data for that determination.  
Given that objects may have exceeded some but not all of the threshold temperatures discussed above, the exact definition of 'primitive' depends on the scientific purpose at hand.
Also, it is incumbent on the mission-design team to 
\revision{quantify the}
risk they are willing to take (note the probabilistic nature of our \revision{temperature} determinations due to the chaotic orbital history of NEOs).


\section{Discussion}
\label{sect:discu}

\subsection{Updated ExploreNEOs thermal-modeling pipeline}

For a fraction of our targets, our data have been published previously along with a straightforward NEATM analysis \citep{Trilling2010,Harris2011}.
Due to the updates in the thermal-modeling pipeline presented in Section~\ref{sect:mc}, our results given in Table \ref{tab:results}
supersede previous values
\revision{where available}.
This difference, however, is always comfortably within the quoted error bars and too small to matter practically. 
\revision{A new analysis of the entire data set including new observations will be presented in a later paper.}

The new pipeline provides  estimates of the statistical uncertainty of diameter and albedo results.  The uncertainties are distributed in a highly non-Gaussian way, especially for \pv, hence asymmetric ``$1\sigma$'' error bars are given.  Additionally, we provide probabilities of \pv\ to fall into specific bins, which allows for more straightforward constraints on the taxonomic type and hence surface mineralogy.

\subsection{Potential spacecraft  targets}

\paragraph{1996 XB27 and 1989 ML:} These 
\revision{bodies}
 are, along with the Hayabusa target Itokawa, the only objects with $\dv < \unit{5}{\kilo\meter\per\second}$ within our sample.  We find both objects to be very high in albedo, indicative of taxonomic types such as E \citep[1989 ML was found to be E type by][]{Mueller2007}.  The chances of them being primitive are very small.
With a diameter of only $\unit{\ensuremath{84^{+13}_{-12}}}{\metre}$, 1996~XB27 is among the smallest celestial objects with known size.


\paragraph{Primitive objects:}

\begin{deluxetable}{lllllllll}
\tablecaption{
\label{tab:primitive} 
Potentially primitive objects as indicated by their low albedo ($p_1 > \unit{50}{\%}$)
}
\tablehead{ \dv & Object & $p_1$ & $D$ &\pv& $T_{10\%}$ & $T_{50\%}$ & $T_{90\%}$ & Taxo\\
(\kilo\metre\per\second) & & & (\kilo\metre)& & (\kelvin) & (\kelvin) & (\kelvin) &}
\startdata 
5.565 & (162998) 2001 SK162 & 0.92 & $ 1.9\pm 0.4$&$ 0.03^{+0.03}_{-0.02}$ & 834 & 462 & 418 & T \\
5.653 & (68372) 2001 PM9    & 0.97 & $ 1.7\pm 0.5$ &  $ 0.018^{+0.017}_{-0.008}$ & 782 & 488 & 441 & -- \\
6.405 & (65679) 1989 UQ &  0.75 & $ 0.7\pm 0.2$ & $ 0.05^{+0.04}_{-0.02}$   & 1082 & 564 & 410 & B\\
6.507 & 1989 AZ         &  0.97 & $ 1.1\pm 0.2$  & $ 0.03^{+0.02}_{-0.01}$  & 861 & 560 & 463 & --\\
6.526 & (100085) 1992 UY4 &0.98 & $ 2.5^{+0.7}_{-0.6}$ & $ 0.023^{+0.019}_{-0.009}$  & 498 & 444 & 417 & P \\
6.609 & (175706) 1996 FG3 &0.84 & $ 1.8^{+0.6}_{-0.5}$ & $ 0.04^{+0.04}_{-0.02}$ & 743 & 565 & 482 & C \\
6.738 & (5645) 1990 SP &   0.60 & $ 2.2^{+0.8}_{-0.7}$ & $ 0.06^{+0.08}_{-0.04}$ & 743 & 540 & 456 & --\\
\enddata
\tablecomments{
See Table \ref{tab:results} for the definition of $p_1$.
Where available, diameter and albedo results are from the reanalysis in
Table \ref{tab:newh}, otherwise from Table \ref{tab:results}.  
As discussed in Section \ref{sect:history}, the chances are \unit{10/50/90}{\%} that the surface temperature has reached $T_{10/50/90\%}$ or above.
}
\end{deluxetable}

Some of our albedo results 
suggest a primitive composition.  We adopt a threshold value of 50\% for $p_1$, the probability of $\pv < \unit{7.5}{\%}$.
A list of all measured physical properties of these objects is compiled in Table \ref{tab:primitive}.
\revision{There are significant object-to-object differences in thermal history.}
As discussed in Section \ref{sect:history}, there is no clearly defined threshold temperature above which primitive material is metamorphosed.
The least heated objects (at the 50\% probability level) are 1992~UY4, 2001~SK162, and 2001~PM9.
1996~FG3 is the only known binary in the low-\pv\ sample.

\section{Conclusions}
\label{sect:conclu}

Of the 293 NEOs observed within the framework of our ongoing ExploreNEOs survey as of 2010 July 14, 
65  have $\dv\leq\unit{7}{\kilo\metre\per\second}$. 
Diameter and albedo measurements for the latter are presented in this work.
Assuming that the rate of observations of low-\dv\ objects stays as it is, 
the number of observed low-\dv\ objects will increase to $\sim 160$ by the end of ExploreNEOs, i.e.\ before the end of 2011.
Teams requiring a physical characterization of potential spacecraft targets are encouraged to contact us.

Out of our 65 low-\dv\ targets, 7 have low albedos indicating a primitive surface composition.  These objects include a binary (1996~FG3) and three objects which stayed remarkably cool during their dynamical history, 
\revision{possibly}
 cool enough to remain primitive: 
1992~UY4, 2001~SK162, and 2001~PM9.

\acknowledgments

Michael Mueller gratefully acknowledges the Henri Poincar\'e Fellowship, which is funded by the
CNRS-INSU and the Conseil G\'en\'eral des Alpes-Maritimes.
The work of MM and MD is supported by ESA grant SSA-NEO-ESA-MEM-017/1.
We thank Patrick Michel for helpful discussions.
This work is based in part on observations made with the Spitzer Space Telescope, which is operated by JPL/Caltech under a contract with NASA. Support for this work was provided by NASA through an award issued by JPL/Caltech.

Facilities: \facility{Spitzer(IRAC)}

\end{document}